\documentclass[preprint,1p]{elsarticle}
\usepackage{lineno}
\usepackage{amsmath,amssymb,graphicx}
\usepackage{amsfonts}
\usepackage{mathrsfs}
\usepackage{epsfig}
\usepackage[active]{srcltx}
\usepackage{cmap}
\usepackage{amsthm}
\usepackage{amsmath}
\usepackage{curves}
\usepackage{amssymb}
\usepackage{mathrsfs}
\usepackage{epsfig}
\usepackage{epstopdf}
\usepackage{graphicx}
\usepackage{subfigure}
\usepackage{amsmath}
\usepackage{amsthm}
\usepackage[dvipsnames]{xcolor}
\usepackage{graphicx}
\usepackage{fancybox}
\usepackage{array}
\usepackage{subfigure}
\numberwithin{equation}{section}
\modulolinenumbers[5]
\usepackage{setspace}
\date{}

\begin{document}

\begin{frontmatter}

\title{The oxygen partial pressure in solid oxide electrolysis cells with two layer electrolytes }

\author[mymainaddress]{Qian Zhang}

\author[mymainaddress]{Qin-Yuan Liu}
\author[mymainaddress]{Beom-Kyeong Park}
\author[mymainaddress]{Scott Barnett}
\author[mymainaddress]{Peter Voorhees}
\address[mymainaddress]{Department of Materials Science and Engineering, Northwestern University}

\begin{abstract}
A number of degradation mechanisms have been observed during the long-term operation of solid oxide electrolysis cells (SOEC).    Using an electrolyte charge carrier transport model, we quantify the oxygen potentials across the electrolyte and thereby provide insights into these degradation mechanisms. Our model describes the transport of charge carriers in the electrolyte when the oxygen partial pressure is extremely low by accounting for the spatial variation of the concentration of oxygen vacancies in the electrolyte. Moreover,  we identify  four quantities that characterize the distribution of oxygen partial pressure in the electrolyte, which are directly related to the degradation mechanisms in the electrolyte as well:   the two oxygen partial pressures at the interfaces of the electrodes and the electrolyte, the oxygen partial pressure at the interface of YSZ/GDC, and the position of the abrupt change in oxygen potential near the p-n junction that develops in YSZ when one side of the cell is exposed to fuel (low oxygen potential, n-type conduction) and the other side is exposed to oxidant (high oxygen potential, p-type conduction).
We give  analytical estimates for all of these quantities. These analytical expressions  provide guidance on the parameters that need to be controlled to suppress the degradation observed  in the electrolyte. In addition, the effects of operating conditions, particularly current
density and operating temperature, on degradation are discussed.

\end{abstract}


\end{frontmatter}

\section{Introduction}
Solid oxide electrolyzer cells, typically consisting of  a YSZ electrolyte, a Ni-YSZ fuel electrode, and a perovskite-based oxygen electrode, have attracted much attention due to their potential to produce hydrogen by H$_2$O electrolysis at a lower cell potential, and hence higher efficiency, than other electrolysis methods \cite{Effe}.  Maintaining a low degradation rate is a key challenge for SOECs. Various degradation phenomena and mechanisms have been reported at high current density \cite{Mogensen4, MChen1}. Oxygen-electrode delamination and pore formation in the YSZ electrolyte near the oxygen electrode have both been reported \cite{Mogensen4,BK} and have been explained by a high oxygen partial pressure at the corresponding position in the electrolyte\cite{Mogensen4,BK,Virkar2}. ZrO2 nanoparticle formation in the Ni-YSZ fuel electrode \cite{MChen1} and delamination within YSZ at extreme low oxygen partial pressure \cite{QY} have been explained by the thermodynamic instability of the Ni-YSZ interface near the fuel electrode. A GDC diffusion barrier layer is typically introduced to prevent reactions between the YSZ
electrolyte and oxygen electrode; cracks and pores have been observed at the GDC/YSZ interface \cite{BK,MChenBi}. It
would be desirable to have a model that can provide a general understanding of these SOEC degradation
phenomena and make quantitative predictions of the conditions expected to cause degradation.

A few different models have been reported.  Oxygen partial pressures at the interfaces between the electrolyte and the two electrodes have been predicted by considering the cell equivalent circuit \cite{Virkar3}, but this approach does not predict the entire oxygen pressure distribution in the electrolyte.  Mogensen and coworkers \cite{Mogensen1,Mogensen2,Mogensen3} proposed a model that accounts for oxygen ion, electron, and hole transport across the electrolyte.  Although this model was successfully used to calculate oxygen potential profiles, further improvements are needed regarding the assumption that oxygen vacancy molar concentration is constant across the electrolyte, and the method for dealing with the interface between YSZ and GDC layers in the electrolyte. Following the pioneering work of Virkar et al and Mogensen et al., their models have been extensivly employed by other researchers \cite{IWC,Virkar} to investigate the oxygen partial pressure distribution in the electrolyte,  in which the general assumption that the conductivity of holes is proportion to $P_{O_2}^{1/4}$ and the conductivity of electrons is proportion to $P_{O_2}^{-1/4}$ was taken based on changing of oxygen vacancy concentration iS negligible. 
Although it is often reasonable to assume that the oxygen vacancy
concentration in YSZ is constant, the variation cannot be neglected at the extremely low oxygen partial
pressure near the fuel side of an electrolysis cell. This is especially true since key degradation phenomena
occur in the electrolyte near the fuel electrode \cite{MChen1,QY} where oxygen partial pressures can be extremely low.
Furthermore, it was assumed the interface between the YSZ and GDC electrolyte layers is abrupt.

In this paper, in the modeling aspect, we propose two improvements upon the prior calculations: (1) allowing the oxygen vacancy molar concentration to vary in YSZ and (2) modeling a two layer YSZ-GDC electrolyte using a diffuse interface model for the two layer YSZ-GDC electrolytes.    In contrast to the sharp interface model introduced previously \cite{Mogensen3}, our diffuse interface approach makes the calculation more straightforward and is more physically realistic since there is generally a thin intermixing layer at the interface of two materials.   The present work focuses on electrolysis operation, due to the extreme oxygen pressures that lead to degradation in this mode; in contrast, oxygen pressure variations in fuel cell operation are relatively small and are not expected to cause degradation.  In recent papers \cite{BK,QY}, the present model were successfully used to quantitatively predict and explain experimental results for delamination of different oxygen electrodes under a range of SOEC operating conditions.  Here we use the model to make more generalized predictions about the effects of current density, temperature and electrolyte and electrode materials, on oxygen pressure and thereby degradation.  

In addition to the numerical results, the main contribution of this paper is that we give analytical estimations to the quantities that characterize the distribution of oxygen partial pressure in the electrolyte, i.e., the oxygen partial pressure at the interface of electrolyte and electrodes, at the interface of YSZ/GDC and the position of abrupt change of oxygen partial pressure in YSZ. This allows experimentalist to  easily estimate  the values of oxygen partial pressures of interest purely based on the operation conditions and measurable quantities in the experiments  without having to perform  simulations.
 
In section 2, the mathematical model is introduced in details. In section 3,  We compare the values of oxygen partial pressure in YSZ and GDC for cases considering changing of molar concentration of oxygen vacancies and cases in which molar concentration of vacancies is taken as a constant. Results and discussion are presented in section 4. 

Finally, we summarize the critical conditions (current density and operating temperature) that lead to degradation.

\section{Mathematical model}

\begin{figure}[htpb]\centering
\subfigure[]{\includegraphics[scale=0.32]{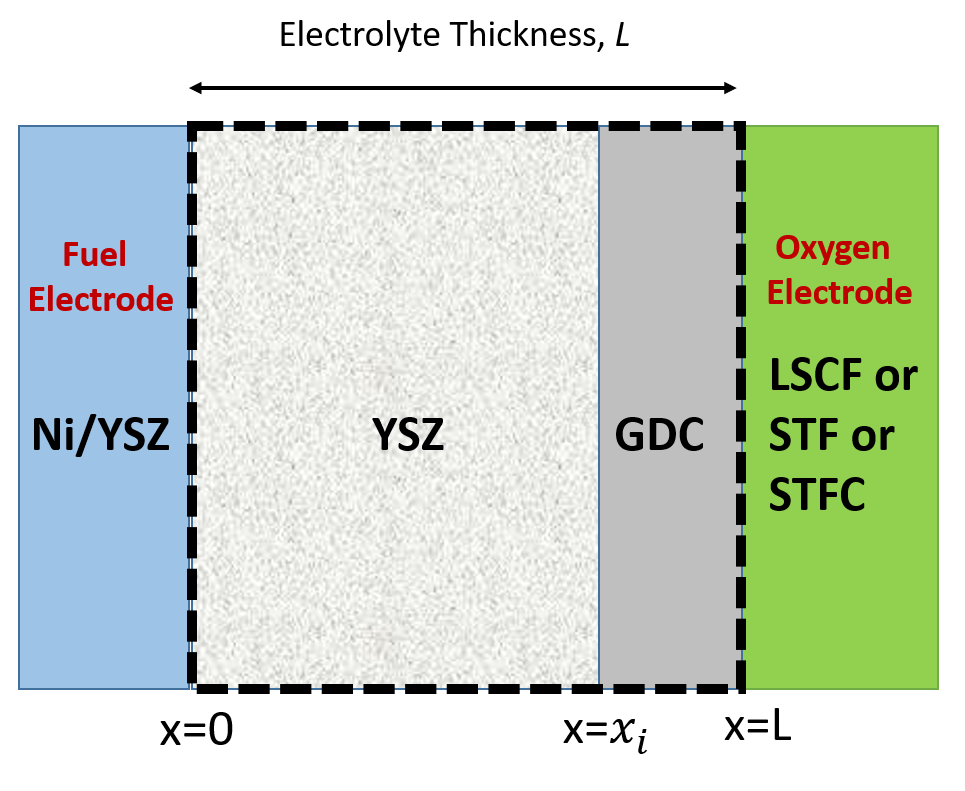}}
\subfigure[]{\includegraphics[scale=0.35]{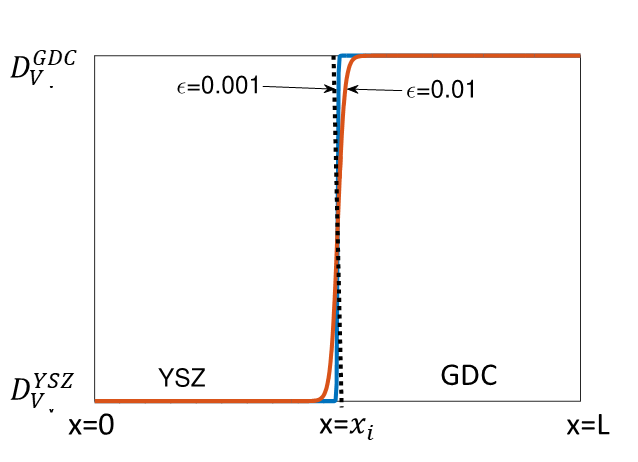}}
\caption{(a) Schematic graph of the cell structure considering in this paper; (b) Schematic graph showing diffusivity of vacancies in the electeolyte as represented by a continuous function $D_V$, which takes the value of $D_V^{YSZ}$ and $D_V^{GDC}$ in the bulk portion of the electrolyte, with a smooth change between the two values in the zone near the interface. The thickness of the interface is proportional to $\epsilon$, see Eq. \eqref{tanh}}.
\label{Diffig}
\end{figure}

The structure of the cell we are considering is shown in Fig. \ref{Diffig}(a), with a domain that is $x\in[0,L]$ along the thickness of the electrolyte from the interface of fuel electrode/electrolyte ($x=0$) to the oxygen electrode/electrolyte ($x=L$). In this paper, unless otherwise specified, YSZ refers to 8YSZ (8 mol$\%$ $Y_2O_3$) and GDC refers to GDC10 (10 mol$\%$ $Gd_2O_3$).  The present model has been verified by the very good agreement obtained when comparing the numerical
and experimental results, see \cite{BK,QY}.

The dominant defects in YSZ and GDC are oxygen vacancies. The reaction represented in Kr\"oger-Vink notation in YSZ and GDC is 
\begin{align}
O_O \rightleftharpoons \frac{1}{2}O_2  +V_O^{\mathbf{\cdot\cdot}}+2e^{'}.\label{reaction1}
\end{align}
Moreover, the electrons can be annihilated and produced through the reaction,
\begin{align}
e^{'}+h^{\mathbf{\cdot}}  \rightleftharpoons  nil.\label{nil1}
\end{align}
Electroneutrality implies the molar concentration of the charge carriers satisfies
\begin{align}
&C_{e}+C_{d}-2C_{V}-C_h=0\label{cn1}
\end{align}
where $C_e$, $C_h$ and $C_V$ are the molar concentration  of electrons, holes and oxygen vacancies, respectively, and $C_d$ is the molar concentrations of the dopant cation, i.e., $C_d$ is the molar concentration of $Y^{'}$ in YSZ and $Gd^{'}$ in GDC. 
The molar concentration of holes and electrons are related by 
\begin{align}\label{Equalibrimnil}
&C_{e}C_h=K_{eh}
\end{align}
where $K_{eh}$ is the equilibrium constant of reaction \eqref{nil1}. This number is a function of the material used for the electrolyte.

At steady state, the current density $i$ in a mixed oxygen conductor is the result of the fluxes of electrons, $j_{e}$, holes, $j_{h}$, and oxygen vacancies $j_{V}$(or ions, $j_{O^{2-}}$, note that $j_{V}$ and $j_{O^{2-}}$ are related due to structure conservation on the anion sub-lattice sites). In the electrolyte,
\begin{align}
i=i_{V}+i_{eh}\quad and \quad i_{eh}=i_e+i_h,
\end{align}
and
\begin{align}
i_s=z_sFj_s,\quad j_s=-\frac{D_sC_s}{RT}\nabla\bar{\mu}_s,\quad {s=V,e,h},
\label{crtoflux}
\end{align}
where $D_s$ is the diffusivity, $z_s$ is the charge number  of the corresponding species. $R$ is the gas constant and $T$ is operating temperature.
Here, the electrochemical potential of $s$ is  
$
\bar{\mu}_s=RT\ln(C_s/C_s^\varnothing)+z_sF\phi,
$
where the supscript $\varnothing$ refers to the corresponding quantity at standard state (for a given temperature, the standard state is taken at a pressure of 1 atm)  and $\phi$ is the Galvani potential. Moreover, $i_{eh}$ and $i_V$ are constants at steady state.

The model is valid for both YSZ and GDC electrolytes.  In the YSZ/GDC two layer electrolyte, the material parameters, such as $D_e$, $D_h$, $D_V$, $C_d$ and $K_{eh}$ are different, which implies that 
these parameters are dependent on the position $x$ in the electrolyte.  We model this by introducing a very thin diffusion layer at the YSZ/GDC interface.  
The diffusivity of oxygen vacancies $D_V$, for example, is given by
\begin{align}
D_V(x)=\frac{D_V^{YSZ}}{2}\left[1+\tanh\left(\frac{x-x_i}{\sqrt{2}\epsilon}\right)\right]+\frac{D_V^{GDC}}{2}\left[1-\tanh\left(\frac{x-x_i}{\sqrt{2}\epsilon}\right)\right],\quad 0\leq x\leq L
\label{tanh}
\end{align}
and shown in Fig. \ref{Diffig}, where the thickness of the interdiffusion layer between YSZ and GDC is given by the parameter $\epsilon$ and $x_0$ is the location of the interface between YSZ and GDC.

The boundary conditions on $C_e$ are determined by the overpotentials ($\eta_H$ and $\eta_O$) at the electrolyte/electrode interface, which can be measured in the experimentally:
\begin{align}
&E_H=E_H^{OCV}+\eta_H=E^\varnothing-\frac{RT}{F}\ln\left(\frac{C^{H}_e}{C_e^\varnothing}\right),\quad &x=0,\label{newfrpa}\\
&E_O=E_O^{OCV}-\eta_O=E^\varnothing-\frac{RT}{F}\ln\left(\frac{C^{O}_e}{C_e^\varnothing}\right),\quad &x=L,
\label{newfrpb}
\end{align}
where $E$ is the chemical potential of electrons in Volts. The subscript and superscript $``H"$ and $``O"$ refer to the quantities at the fuel electrode/YSZ electrolyte and oxygen electrode/GDC electrolyte interfaces, respectively. $E_{H(O)}^{OCV}$, the chemical potential of electrons in open circuit voltage case, that can be obtained by the equilibrium reaction \eqref{reaction1} from the different  oxygen partial pressures at the two electrodes. Using the boundary conditions (Eq. \eqref{newfrpa} and Eq. \eqref{newfrpb}) and equations  Eq. \eqref{cn1}-\eqref{crtoflux}, we can determine  $C_e$ and $\phi$, as shown in S.1.

It is possible to use different methods of calculating the overpotentials  in the model. For simplicity, the overpotentials $\eta_{H(O)}$ here are estimated from the measured electrode polarization resistances $R_P^{H(O)}$ values using an approximate form of the Butler-Volmer equation \cite{BK,Justin}:
\begin{align}
\eta_{H(O)}=\frac{RT}{F}\sinh^{-1}({iR_P^{H(O)}F}/{RT}).
\label{BVE}
\end{align}

\begin{figure}[htpb]\centering
\subfigure[YSZ]{\includegraphics[scale=0.27]{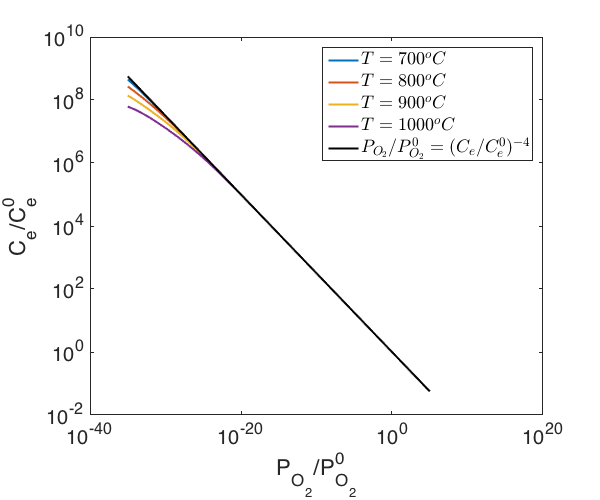}}
\subfigure[GDC]{\includegraphics[scale=0.27]{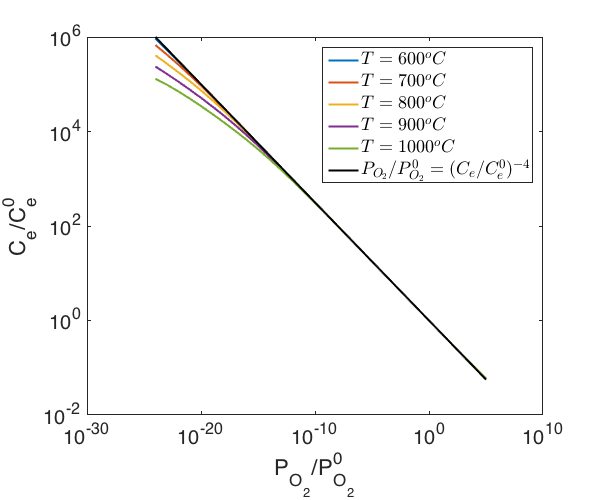}}
\caption{Comparasion the relation between oxygen partial pressure and molar concentration of electrons in YSZ and GDC for cases considering changing of molar concentration of oxygen vacancies and cases, in which $\mu_{V}=\mu_{V}^\varnothing$ (black lines in the figures). In the figures, $C_e^0=C_e^\varnothing$ and $P_{O_2}^0=P_{O_2}^\varnothing$. The lines other than black line are obtained in Eq. \eqref{changecv} at different temperatures. }
\label{PO2VSCE}
\end{figure}

In addition, at equilibrium, and using the expressions for the chemical potentials of oxygen, vacancies and oxygen ions, the reaction \eqref{reaction1} implies: 
\begin{align}
P_{O_2}=P_{O_2}^\varnothing\left(\frac{C_e}{C_e^\varnothing}\right)^{-4}\left(\frac{C_V}{C_V^\varnothing}\right)^{-2}\left(\frac{C_{O_O}}{C_{O_O}^\varnothing}\right)^{2}.
\label{changecv}
\end{align}
Here, $C_V$ can be obtained by the electroneutrality condition Eq. \eqref{cn1} and the local equilibrium of the reaction shown in  Eq.\eqref{Equalibrimnil}, 
\begin{align}
C_V=\frac{1}{2}(C_e+C_d-\frac{K_{eh}}{C_e}).
\end{align}
The molar concentration of oxygen ions $C_{O_O}$ can be obtained using site conservation: 
$
C_{O_O}=C_{anion}-C_V,
$
where $C_{anion}$ is the molar concentration of anion sublattice sites. Thus, we can determine the value of the oxygen partial pressure $P_{O_2}$, once we know  $C_e$. It is obvious that $P_{O_2}\propto C_e^{-4}$, if the change of concentration of vacancies can be neglected in the electrolyte (see also  the black solid line in Fig. \ref{PO2VSCE}). Note that in the electrolyte, $P_{O_2}$ is introduced as a measure of the chemical potential of oxygen molecules in the electrolyte, it is not necessary to generate a gas phase in the electrolyte.

At the hydrogen electrode, the concentration of electrons under open circuit conditions ($C_e^{H,OCV}$) can be determined using Eq. \eqref{changecv} for a given ambient oxygen partial pressure ($P_{O_2}^{H,amb}$) at the fuel electrode. If the ambient oxygen partial pressure at the fuel electrode is not extremely low (see Fig. \ref{PO2VSCE}(a)), $C_e^{H,OCV}$ can be estimated as $$C_e^{H,OCV}=C_e^{\varnothing,YSZ} \left({P_{O_2}^{H,amb}}\right)^{-\frac{1}{4}},$$ by the equilibrium of reaction Eq. \eqref{reaction1}.
Then, according to Eq. \eqref{newfrpa} and Eq. \eqref{BVE}, the concentration of electron at the interface of hydrogen electrode and electrolyte is
\begin{align}\label{CEH}
   {C_e^H}={C_e^{H,OCV}}\exp{[-\sinh^{-1}(iR_P^HF/RT)]}.
\end{align}
Thus, the oxygen partial pressure at the interface of hydrogen electrode and electrolyte can be obtained by substituting Eq.\eqref{CEH} into Eq. \eqref{changecv}, i.e., 
\begin{align}\label{po2H}
    P_{O_2}^H=P_{O_2}^\varnothing\left(\frac{C_e^H}{C_e^{\varnothing,YSZ}}\right)^{-4}\left(\frac{C_V^H}{C_V^{\varnothing,YSZ}}\right)^{-2}\left(\frac{C_{O_O}^H}{C_{O_O}^{\varnothing,YSZ}}\right)^{2}.
\end{align}

At the oxygen electrode, the concentration of electrons at open circuit  ($C_e^{O,OCV}$) in the high oxygen partial pressure ($P_{O_2}^{O,amb}$) ambient environment, can be estimated as
\begin{align}
    C_e^{O,OCV}=C_e^{\varnothing,GDC}\left({P_{O_2}^{O,amb}}\right)^{-\frac{1}{4}}.
\end{align}
Then according to Eq. \eqref{newfrpb} and Eq. \eqref{BVE}, the concentration of electrons at the interface of oxygen electrode and electrolyte is
\begin{align}
    C_e^{O}=C_e^{\varnothing,GDC}\left({P_{O_2}^{O,amb}}\right)^{-\frac{1}{4}}\exp{\left[\sinh^{-1}\left(\frac{iR_P^OF}{RT}\right)\right]}.
\end{align}
Thus, the oxygen partial pressure at the interface of oxygen electrode and electrolyte can be estimated as
\begin{align}\label{po2O}
    P_{O_2}^O=P_{O_2}^{O,amb}\exp[-4\sinh(iR_P^OF/RT)].
\end{align}


\subsection{Degradation phenomena discussed in this paper}
 At high current densities and low temperatures, severe degradation has been observed \cite{BK,MChenBi,QY}. In the following, we summarize known degradation phenomena (C1-C3) due to the distribution of oxygen partial pressure in two-layer YSZ/GDC electrolyte:

{\it{C1. Delamination at the interface of the electrolyte and oxygen electrode:}}  {When
the cells are operating under the electrolysis mode, the oxygen potential or the
oxygen equilibrium pressure increases inside the electrolyte adjacent to the
oxygen electrode. If sufficient pressure builds up in pre-existing flaws or
crack, then cracks can grow. In \cite{Virkar2}, Virkar gives a threshold of the oxygen partial pressure ($P_{O_2}^{O-cr}$) for the crack growth: 
\begin{align}
\displaystyle P_{O_2}^{O-cr}=\frac{1}{2}\sqrt{\frac{\pi}{(1-v^2)c}}K_{IC}
\label{cripo2}
\end{align}
where $v$ is Poisson's ratio and $K_{IC}$ is the fracture
toughness of the material, c is the size of preexisting pore.
For instance, if the crack size (c) is $1\mu m$, $v = 0.24$ and $K_{IC} = 0.8MPa\sqrt{m}$ for perovskite, the critical pressure ($P_{O_2}^{O-cr}$) is about $7\times 10^3atm$. }

{\it{C2. Thermodynamic instability of Ni-YSZ at extremely low oxygen partial pressure close to the interface of fuel electrode and electrolyte:}}  Intermetallic phases Ni$_{x}$Zr$_{y}$ may form at the interface under low oxygen activity. The critical oxygen partial pressure ($P_{O_2}^{H-cr}$) for the formation of a Ni-Zr compound is $3\times 10^{-29\pm 3} atm$ at 850$^oC$ \cite{MChen1}. According to the data in \cite{MChen1}, the Gibbs energy of reaction for Zr reduction into Ni-Zr compound is esitmated to be $600kJ/mol$. As claimed in \cite{MChen1}, ``A difference of about $50 kJ/mol$ in the Gibbs energy is within a normal uncertainty range for measured enthalpy.''  Thus, in this paper, we estimate the critical $P_{O_2}^{H-cr}$ of getting reduction at temperature $T(K)$ is between 
$\exp[(600+54.5)kJ/RT]$ and $\exp[(600-54.5)kJ/RT]$, which can result in a difference of 3 decades in the calculated critical $P_{O_2}$.

{\it{C3. Crack or pore formation at the YSZ/GDC interface:}} Cracks and pores have been observed to form at the YSZ/GDC interface \cite{BK}, which is believed to be associated with the high oxygen partial pressure at this interface. Here, we estimate the critical oxygen partial pressure ($P_{O_2}^{I-cr}$) for crack formation at YSZ/GDC interface using Eq. \eqref{cripo2} for YSZ.
\section{The effect of a varying oxygen vacancy concentration}
Figure \ref{PO2VSCE} shows the relationship between the oxygen partial pressure $P_{O_2}$ and the molar concentration of electrons $C_e$ in YSZ and GDC for two cases – when the molar concentration of oxygen vacancies $C_V$ is allowed
to vary and constrained to be constant. At higher $P_{O_2}$ environment, allowing $C_V$ to change or be a constant
yields almost identical values. At lower $P_{O_2}$, there is a significant difference because, in order to reach
equilibrium, a low $P_{O_2}$ will drive the reaction shown in \eqref{reaction1} from the left side to the right side, which
introduces a change in $C_V$ and $C_e$. {Higher temperatures with low oxygen partial pressure shifting the equilibrium more to the right hand side of reaction  \eqref{reaction1}.} The results are
consistent with the experimental measurements in Ref. \cite{deltapo2}.

\section{Results}
In the following, in order to better address the experiments,  we investigate 
(a) the distribution of $P_{O_2}$ in the YSZ/GDC bilayer electrolyte; 
(b) the influence of operating conditions (current density $i$ and operating temperature $T$); (c) the influence of transport properties of the electrolyte on the distribution of $P_{O_2}$ in the YSZ/GDC electrolyte. Parameters regarding transport properties of YSZ and GDC are summarized in S.4. Other parameters taken in the numerical simulations are shown in the captions of the figures.
\subsection{Distribution of oxygen partial pressure in the YSZ/GDC bilayer electrolyte:}

\begin{figure}[htpb]\centering
\subfigure[]{\includegraphics[scale=0.3]{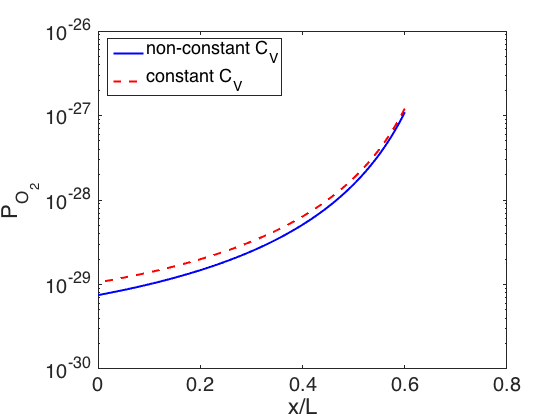}\label{PortionRs}}
\subfigure[]{\includegraphics[scale=0.3]{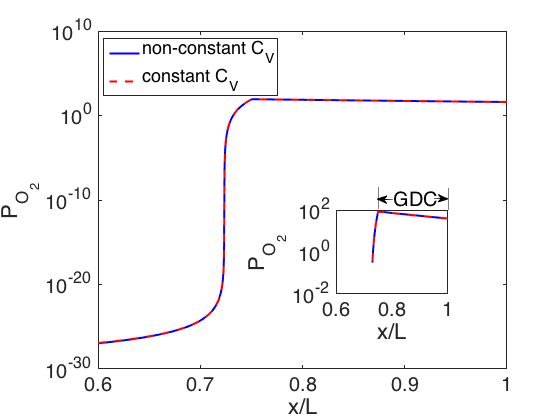}\label{SOECPTPO2}}
\caption{Distribution of oxygen partial pressure in the electrolyte for full cell operating in the SOEC mode assuming a constant vacancy concentration \cite{Mogensen3}, and non-constant vacancy concentration, (a) along the part of electrolyte close to the fuel electrode and (b) along the part of electrolyte close to the oxygen electrode. The zoom-in figure shows the peak pressure at the interface of YSZ/GDC. Parameters in the numerical simulations are taken as, $i=-1.6A/cm^2$, $T=800^oC$, $L=50\mu m$, $L_{GDC}/L=0.25$ and $R_P^H=0.9\Omega/cm^2$, $R_P^O=0.1\Omega/cm^2$, $P_{O_2}^{H-OCV}=10^{-23}(atm)$ and $P_{O_2}^{H-OCV}=0.2(atm)$. In the figures, $0\leq x/L<0.75$ is YSZ. $0.75<x/L\leq 1$ is GDC.}
\end{figure}

Figure \ref{SOECPTPO2} shows the same basic variation in $P_{O_2}$ with electrolyte position under the SOEC operating mode ($i<0$), that has been reported previously \cite{Mogensen3} when a GDC layer is present. As discussed later, there is an abrupt
change from high to low $P_{O_2}$ in the electrolyte. A $P_{O_2}$ maximum is predicted at the interface of YSZ and GDC,
see Fig. \ref{PortionRs}. This is the same as what has been found in \cite{Mogensen3} on the high $P_{O_2}$ side where $C_V$ is constant. As
shown in Fig. \ref{PortionRs}, however, on the fuel side of the cell $P_{O_2}$ is significantly lower when $C_V$ is constant versus
being allowed to vary. 

In addition, as shown in Fig. \ref{PortionRs} and \ref{SOECPTPO2}, the distribution of oxygen partial pressure in the electrolyte is a sigmoid shaped curve with a maximum value at the interface of YSZ/GDC. It is easy to see that the oxygen partial pressure at the interface of electrolyte and electrodes(at $x=0$ and $x=1$), at the interface of YSZ/GDC and the position of the abrupt change of value of oxygen partial pressure in YSZ characterize the distribution of oxygen partial pressure in the electrolyte.   In the following, we give analytical estimate ofo these quantities based on the model we proposed in Sec.1.

\subsection{Estimate of the oxygen partial pressure ($P_{O_2}^i$) at the interface of YSZ/GDC:}
In the following, we investigate the increase in the oxygen partial pressure from GDC/oxygen electrode to YSZ/GDC interface. At steady state, $i_V$ and $i_{eh}$ are constants in the electrolyte. Thus, at the interface of YSZ and GDC,  $i_{eh}^{YSZ}=i_{eh}^{GDC}$ and $i_V^{YSZ}=i_V^{GDC}$, where the superscripts refer to the quantities in the corresponding materials.
In particular, at the interface of YSZ and GDC, both YSZ and GDC are p-type conductors, and thus the current density of electrons and holes ($i_{eh}$) in the YSZ and GDC is dominanted by conduction hole ($i_{eh}\approx i_h$), which implies that $i_h^{YSZ}\approx i_h^{GDC}\mbox{ at the YSZ/GDC interface}$. Using equation Eq. \eqref{crtoflux}, $i_h^{YSZ}\approx i_h^{GDC}$ implies \begin{align}\sigma_h^{YSZ}(\nabla\mu_h^{i,YSZ}/F+\nabla\phi^{i,YSZ})
\approx\sigma_h^{GDC}(\nabla\mu_h^{i,GDC}/F+\nabla\phi^{i,GDC}),
\label{conieh}
\end{align}
at the YSZ/GDC interface. The superscript $i$ in Eq. \eqref{conieh} and other equations in this section refers to their values at the interface of YSZ/GDC.  Here, the conductivity $\sigma_s=D_sC_sz_s^2F^2/RT,\quad s=V,e,h,$.  Thus, the driving force for the motion of holes has two components. One  is the gradient of the Galvani potential ($\nabla\phi$) in the electrolyte, and other is the gradient of the chemical potential ($\nabla\mu_h$) in the electrolyte. 


At steady state, $i_{V}$ is a constant across the electrolyte (i.e., $i_{V}^{YSZ}=i_{V}^{GDC}$). 
In addition, at high oxygen partial pressure, the molar concentration of oxygen vacancies can be taken as constant as shown in our previous discussion in Section 3.  Thus, the balance of $i_V$ implies that
%
%
%
\begin{align}
i_V\approx-\sigma_{V}^{YSZ}\nabla\phi^{i,YSZ}\approx-\sigma_V^{GDC}\nabla\phi^{i,GDC}\label{EQIV}.
\end{align}
 
Therefore, substituting Eq. \eqref{EQIV} into Eq. \eqref{conieh} implies 
\begin{align}\sigma_h^{GDC}\nabla\mu_{h}^{i,GDC}\approx{\sigma_h^{YSZ}}\nabla\mu_h^{i,YSZ}-{i_V}F\left(\frac{\sigma_h^{YSZ}}{\sigma_V^{YSZ}}-\frac{\sigma_h^{GDC}}{\sigma_V^{GDC}}\right).
\label{ECR1}
\end{align}
Inspired by our numerical simulation results, the first term (${\sigma_h^{YSZ}}\nabla\mu_h^{i,YSZ}$) at the right hand side of the above equation is negligible comparing with the second term at the right hand side and the term at the left hand side of the above equation. Thus, the sign of $\nabla\mu_h^{i,GDC}$ is determined by the sign of 
\begin{align}
\frac{\sigma_h^{YSZ}}{\sigma_V^{YSZ}}-\frac{\sigma_h^{GDC}}{\sigma_V^{GDC}}\label{DV}
\end{align} under SOEC mode ($i<0$).  According to the value of parameters taken in S.4, the sign of Eq. \eqref{DV} is negative for operating temperatures from $600^oC$ to $1000^oC$. This implies $\nabla\mu_h^{i,GDC}<0$, thus $\nabla C_h^{i,GDC}<0$, which implies $\nabla C_e^{i,GDC}>0$ because $C_h^{i,GDC}C_e^{i,GDC}=K_{eh}^{GDC}$. According to Fig. \eqref{PO2VSCE}, at high oxygen partial pressure, $P_{O_2}\propto C_e^{-4}$ in both YSZ and GDC, and thus we have $\nabla P_{O_2}^{i,GDC}<0$. Moreover, due to the results in \cite{Mogensen1}, at high oxygen partial pressure,  $\nabla C_e^{i,YSZ}<0$ and $\nabla P_{O_2}^{i,YSZ}>0$.  Since the gradient in $P_{O_2}$ changes at the YSZ/GDC interface, there must be a maximum in the distribution of oxygen partial pressure at the interface of YSZ and GDC layer. This is different than the discussion in \cite{Mogensen3} where they propose that the maximum is introduced because of difference of the electron conductivity between YSZ and GDC. Here, we show that the electron conductivity does not contribute significantly on the oxygen partial pressure at the interface of YSZ and GDC because it is hole conductor ($i_{eh}\approx i_h$) at this location.

In the following, we give estimation of the value of oxygen partial pressure ($P_{O_2}^i$) at the interface of YSZ/GDC. Our estimation can help experimentalists with an evaluation on the value of oxygen partial pressure at the interface of YSZ/GDC without solving the complicated mathematical model.

Inspired by our numerical results, where we find that the variation of $\mu_h^{GDC}$ with position is approximately linear  within GDC layer. Thus, we can estimate $\nabla\mu_h^{GDC}$ with $\nabla\mu_h^{i,GDC}$, integrate Eq. \eqref{ECR1} over the thickness of GDC layer, 
\begin{align}\label{EqRB}
\frac{RT}{L_{GDC}}\left[\ln \left(\frac{C_e}{C_e^{\varnothing}}\right)^i-\ln \left(\frac{C_e}{C_e^\varnothing}\right)^O\right]\approx -\frac{i_VF}{\sigma_h^{GDC}}\left(\frac{\sigma_h^{YSZ}}{\sigma_V^{YSZ}}-\frac{\sigma_h^{GDC}}{\sigma_V^{GDC}}\right)
\end{align}
where the `$i$'  and `$O$' superscript represent the quantities at the interface of YSZ/GDC and GDC/oxygen electrode respectively. In the high oxygen partial pressure environment, the equilibrium of reaction Eq. \eqref{reaction1} implies
\begin{align}\label{EqR1}
-\frac{\mu_e^\varnothing}{F}-\frac{RT}{F}\ln\frac{C_e}{C_e^\varnothing}=\frac{\mu_{O_2}^\varnothing+2\mu_{V_O^{\cdot\cdot}}^\varnothing-2\mu_{O_O}^\varnothing}{4F}+\frac{RT}{4F}\ln\frac{P_{O_2}}{P_{O_2}^\varnothing}
\end{align}
in which $E^\varnothing=-{\mu_e^\varnothing}/{F}$. According to equilibrium condition of reaction Eq. \eqref{reaction1}, $$-\frac{\mu_e^\varnothing}{F}=\frac{\mu_{O_2}^\varnothing+2\mu_{V_O^{\cdot\cdot}}^\varnothing-2\mu_{O_O}^\varnothing}{4F}.$$

Substituting Eq. \eqref{EqR1} into left hand side of Eq. \eqref{EqRB} implies,
\begin{align}\label{AnalyticalPOI}
\ln P_{O_2}^i-\ln P_{O_2}^O\approx  \frac{4iFL_{GDC}}{RT\sigma_h^{GDC}}\left(\frac{\sigma_h^{YSZ}}{\sigma_V^{YSZ}}-\frac{\sigma_h^{GDC}}{\sigma_V^{GDC}}\right),
\end{align}
where we replace $i_V$ with $i$ for leakage current density is small.
This gives us a quantitative estimation on the value of oxygen partial pressure at the interface of the YSZ/GDC.

According to our discussion in section 3, substitute Eq. \eqref{po2O} into Eq. \eqref{AnalyticalPOI}, the oxygen partial pressure at the interface of YSZ/GDC is,
\begin{align}
P_{O_2}^i\approx &P_{O_2}^{O,amb}\exp\left[-
\frac{16iFL_{{GDC}}}{RT\sigma_h^{{GDC}}}\left(\frac{\sigma_h^{YSZ}}{\sigma_V^{YSZ}}-\frac{\sigma_h^{GDC}}{\sigma_V^{GDC}}\right)\sinh^{-1}\left(\frac{iR_P^OF}{RT}\right)\right],
\label{EstiPO2}
\end{align}
where $P_{O_2}^{O,amb}$ is the ambient oxygen partial pressure at the oxygen electrode, which is usually taken as 0.2 atm. Note that,  $\sigma_h^{YSZ}$ and $\sigma_h^{GDC}$ are proportional to $P_{O_2}^{1/4}$ because $C_h\propto P_{O_2}^{1/4}$ at high oxygen partial pressure. However, such a correlation can be eliminated between the denominator ($\sigma_h^{GDC}$) and numerator($\sigma_{h}^{GDC}$ and $\sigma_h^{YSZ}$) in Eq. \eqref{EstiPO2}. In addition, according to our discussion in section 3, at high oxygen partial pressure, molar concentration of vacancies can be taken as constant. Thus, the right hand side of Eq. \eqref{EstiPO2} is \textit{independent} of $P_{O_2}^i$. 

The above estimate is also verified by the numerical simulations shown in Fig.\ref{YSZlayer} and Fig. \ref{GDClayer}. This shows that the above analysis gives an accurate prediction on the magnitude of oxygen partial pressure at the interface of YSZ/GDC relative to the oxygen partial pressure. 

According to the estimate of on the oxygen partial pressure obtained in Eq. \eqref{EstiPO2}, it shows that the peak pressure at the interface of YSZ/GDC is introduced because the difference of the ratio of conductivity of holes and oxygen vacancies in YSZ and GDC. Moreover, the magnitude of the peak pressure at the interface of YSZ/GDC under SOEC mode can be reduced by reducing the GDC layer thickness.  Changes in operating conditions, i.e.,  lowering the current density and increasing operating temperature can also reduce the peak pressure.

\subsubsection{Effect of conductivity of holes and vacancies:}   
\begin{figure}
\subfigure[]{\includegraphics[scale=0.25]{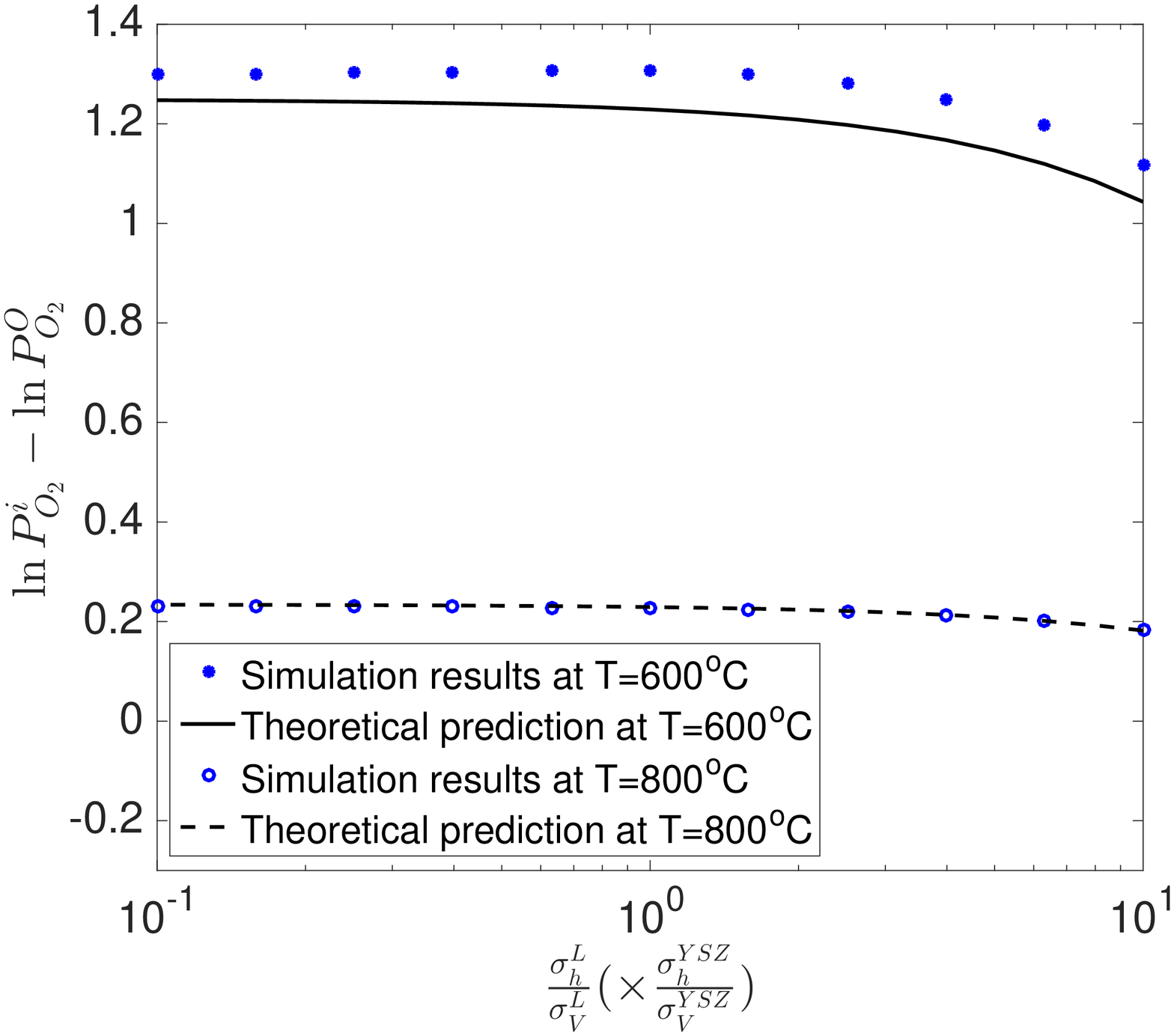}\label{YSZlayer}}
\subfigure[]{\includegraphics[scale=0.25]{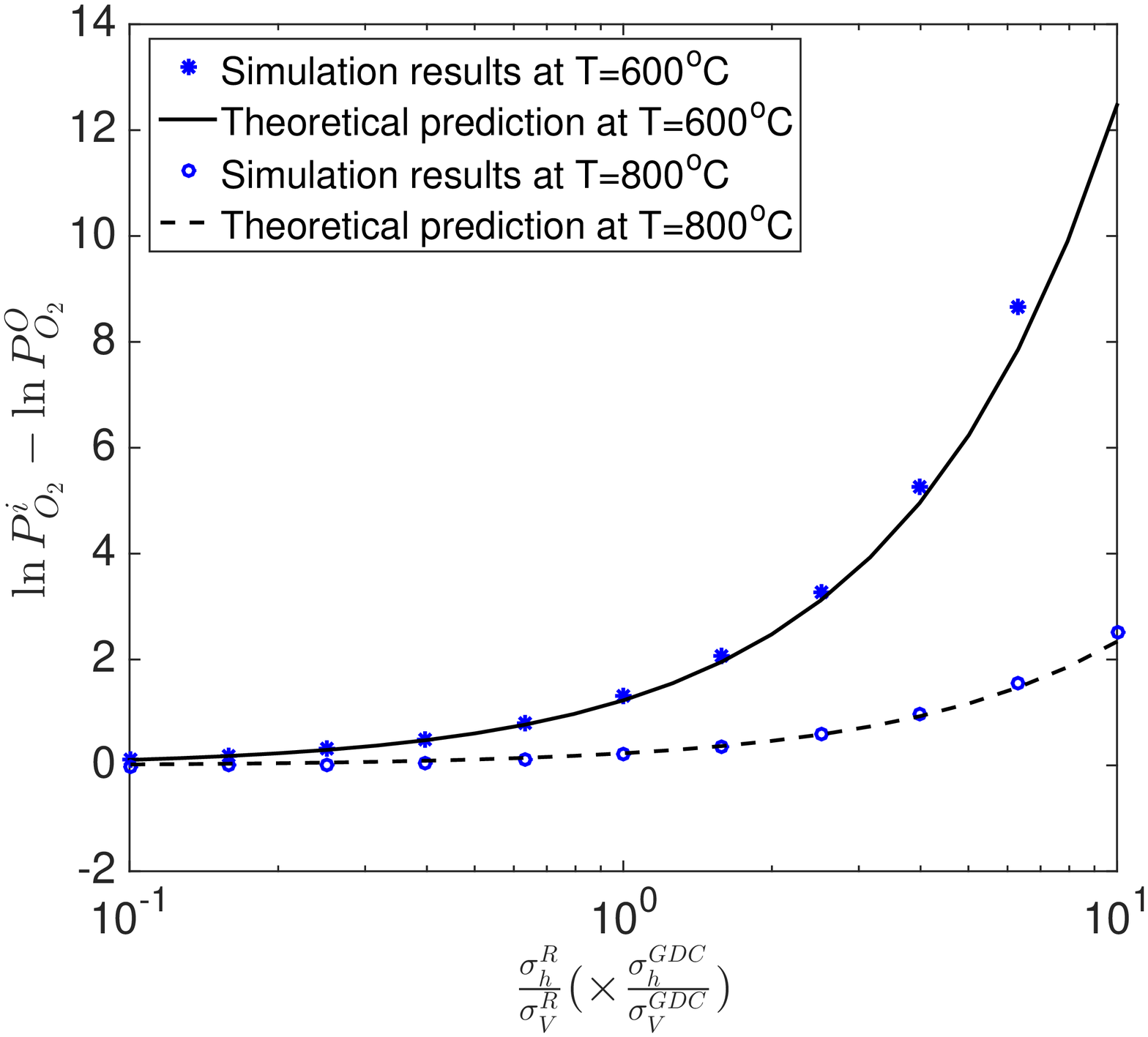}\label{GDClayer}}
\caption{Difference of oxygen partial pressure at the interface of YSZ/GDC ($P_{O_2}^i$) and the oxygen electrode/GDC electrolyte ($P_{O_2}^O$) changes with $\sigma_h/\sigma_V$ varying by 10 times in (a) YSZ layer (relative to the reference value of YSZ) and (b) GDC layer (relative to the reference value of GDC). Other parameters taken in the numerical simulations are $i=-0.8A/cm^2$, $L=20\mu m$ and $L_{GDC}/L=0.25$, $R_P^H=R_P^O=0.1 \Omega\cdot cm^2$, $H_2:H_2O=97\%:3\%$ at fuel electrode and $P_{O_2}^{H-OCV}=0.2(atm)$.}
\end{figure}

The conductivity ratio of holes and oxygen vacancies $\sigma_h /\sigma_V$ can be varied by using different electrolyte compositions, e.g., by changing the Gd doping concentration in GDC or the Y doping concentration in YSZ. Fig. \ref{YSZlayer} shows an increase in the excess $P_{O_2}$ at the GDC/YSZ interface by varying the diffusivity of oxygen
vacancies $D_V$ by factors from 0.1 to 10 relative to the reference value of the 8YSZ electrolyte (on the low-$P_{O_2}$ side of the electrolyte). Fig. \ref{GDClayer} shows an decrease in the excess $P_{O_2}$ at the GDC/YSZ interface by varying the diffusivity of oxygen
vacancies $D_V$ by factors from 0.1 to 10 relative to the reference value of the GDC10 electrolyte (on the high-$P_{O_2}$ side of the electrolyte). Moreover, we obtain
very good agreement between the numerical results and our theoretical prediction as shown in Fig. \ref{YSZlayer} and
\ref{GDClayer}. In summary, Figs. \ref{YSZlayer} and \ref{GDClayer} show that increasing of $\sigma_h /\sigma_V$ in the layer close to the fuel electrode and
decreasing of $\sigma_h /\sigma_V$ in the layer close to the oxygen electrode reduces the magnitude of oxygen partial
pressure at the interface of YSZ/GDC. In experiments, this can be realized by replacing 8YSZ by 15YSZ and/or
replacing GDC10 by GDC15, as shown in Fig. \ref{YSZlayer} and \ref{GDClayer}.

\subsection{Position of inflection point in $P_{O_2}$ in the YSZ electrolyte}
As shown in Fig. \ref{SOECPTPO2}, there is an abrupt change in the oxygen partial pressure in the YSZ electrolyte, that leads to an inflection point in the spatial variation in the oxygen partial pressure.  Together with the estimate of the oxygen partial pressure at the interface of YSZ/GDC obtained in Eq. \eqref{EstiPO2} and the oxygen partial pressure at the interface of electrolyte and electrodes, the inflection point in YSZ part of the electrolyte characterizes the distribution of oxygen partial pressure in the two-layer YSZ/GDC electrolyte. Moreover, experimental results show \cite{dong} the abrupt change of grain size in the
electrolyte can be coincident with the inflection point of the distribution of oxygen partial pressure in the electrolyte.
This suggests that the distribution of oxygen partial pressure can be an indicator of the microstructure
in YSZ,  which can then influence the conductivity of oxygen ion and the electrons
in the electrolyte. 

For simplicity, in the following, we only consider a single layer YSZ electrolyte when one side of the cell is exposed to fuel and the other side is exposed to oxidant.

\begin{figure}[htpb]\centering
\subfigure[]{\includegraphics[scale=0.4]{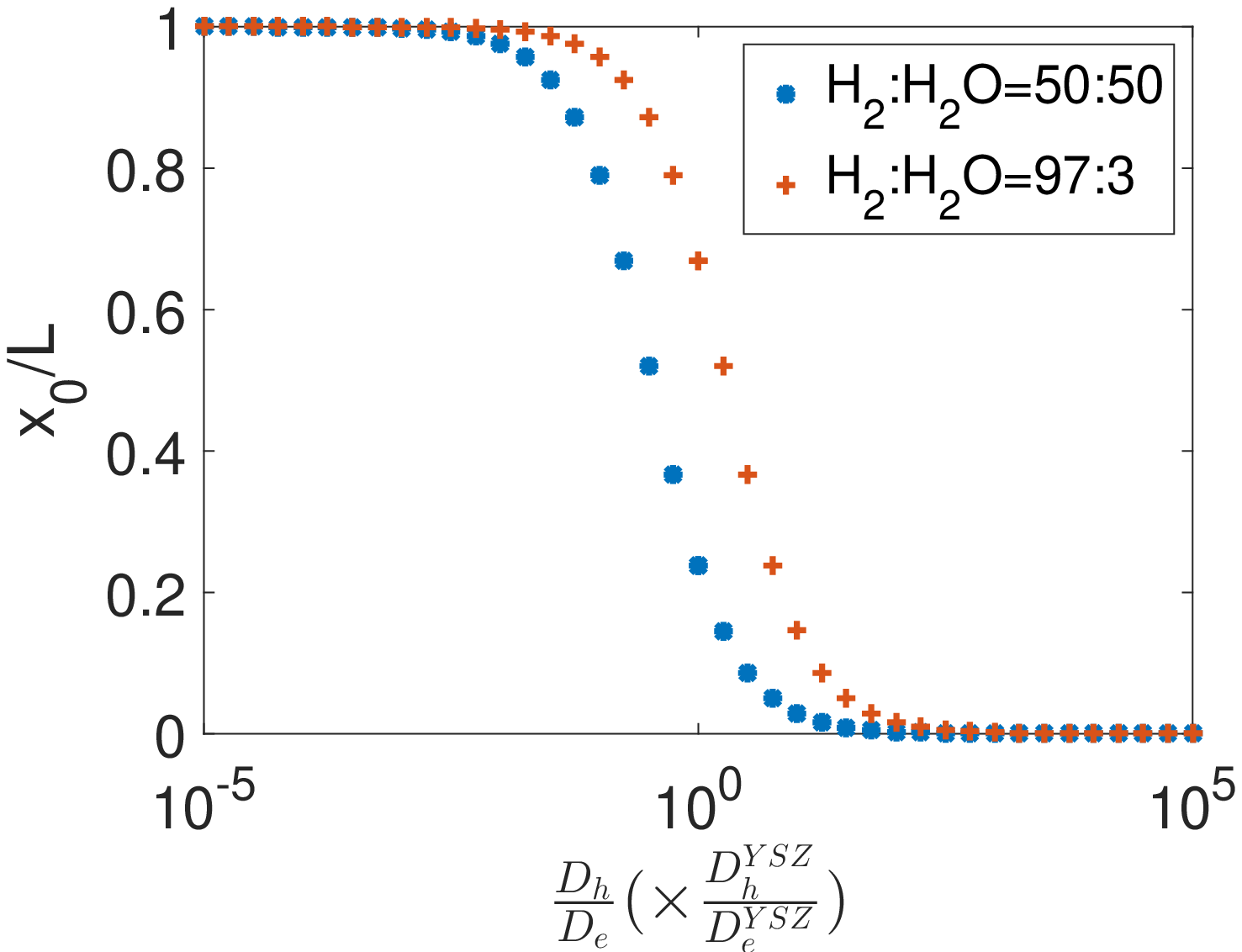}\label{inflenA}}
\subfigure[]{\includegraphics[scale=0.4]{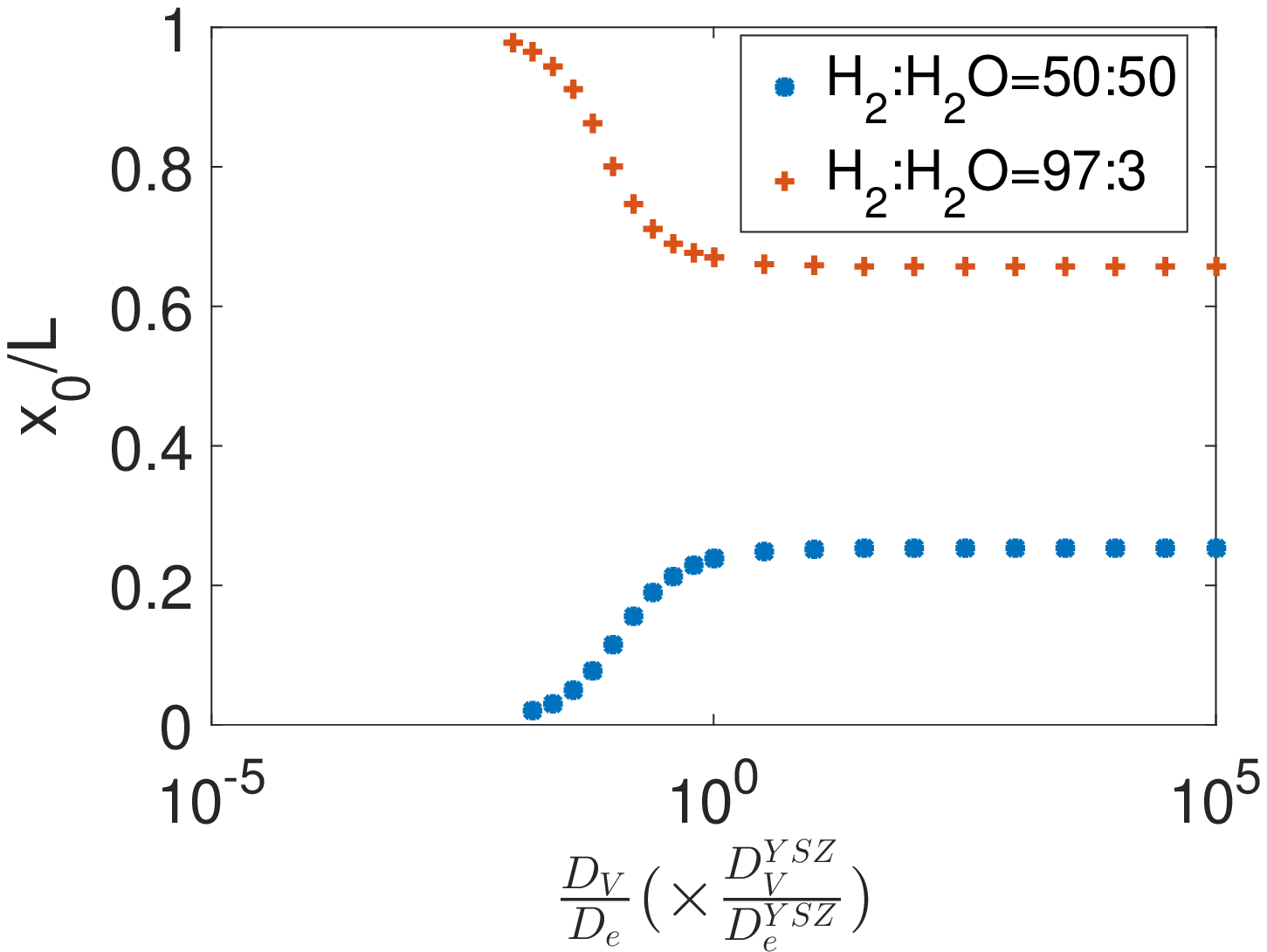}\label{inflenCoverB}}
\caption{ The position of the inflection point in $\log P_{O_2}$ in the electrolyte as (a) a function of $D_h/D_e$ and (b) as a function of $D_V/D_e$. Other parameters in the numerical simulations are taken as $i=-0.8A/cm^2$, $T=800^oC$ and $L=12.5\mu m$, $R_P^H=R_P^O=0.1\Omega\cdot cm^2$.}
\end{figure}

 If we scale the transport properties of the electrolyte by the diffusivity of electrons ($D_e$), there are two parameters that characterize
the properties of the electrolyte that influences the  spatial variation in $\ln P_{O_2}$ , the first is the ratio of the diffusivity  of holes to that of the
electrons in the electrolyte. Due to variations in the synthesis process, $D_h/D_e$ may vary in the same kind of material. 
As $D_h/D_e$ increases, the inflection point moves from the $O-$side to $H-$side, see Fig. \ref{inflenA}. 

The second parameter that characterizes the electrochemical properties of the electrolyte is the ratio of fluxes of oxygen ions and electrons. 
As ${D_V}/{D_e}$ is increased, the current density of oxygen ions approaches the applied current density (i.e., $ i_{V}\approx i$ and $\nabla\phi\approx 0$) and the relative position of the inflection point ($x_0/L$) converges to a constant (see Fig. \ref{inflenCoverB}). Analytical analysis suggests (see, S.2) that the approximate position of the inflection point  is given by
\begin{align}\label{inflectionlimit}
\displaystyle\frac{x_0}{L}=\frac{\left[\displaystyle\frac{D_h^{YSZ} K^{YSZ}_{eh}}{D_e^{YSZ}(C_e^\varnothing)^2}-C_e(0)^2\right]C_e(1)}{\left[C_e(1)-C_e(0)\right]\left[\displaystyle\frac{D_h^{YSZ}K_{eh}^{YSZ}}{D_e^{YSZ}(C_e^\varnothing)^2}+C_e(0)C_e(1)\right]}.
\end{align}
This estimate is consistent with our numerical results.
 Another interesting result is that the position of the inflection point moves from the H-side towards the O-side with increasing of $D_V/D_e$, when the concentration portion of $H_2$ is $50\%$ at the $H_2$ electrode (see the blue line in Fig. \ref{inflenCoverB}). By contrast, the position of the inflection point moves from the O-side to the H-side with the increasing $D_V/D_e$, when the gas on the fuel side changes to $97\%$ $H_2$ (see the red dots in Fig. \ref{inflenCoverB}). That is, the changing direction of 
locations of the inflection point  varies with the $H_2$ concentration at the fuel side.

Moreover, as shown in Fig. \ref{inflenCoverB}, we find that Eq. \eqref{inflectionlimit} also gives a good approximation of the position of inflection point of the oxygen partial pressure distribution in YSZ.  In the case of typical YSZ8 where $D_V/D_e=D_V^{YSZ}/D_e^{YSZ}$ with gas composition at the fuel side $H_2:H_2O=97:3$, the position of inflection point is $x_0/L=0.6690$ whereas the value given by Eq. \eqref{inflectionlimit} is $x_0/L=0.6565$. For the case with $H_2:H_2O=50:50$, the position of the inflection point is $x_0/L=0.2377$, wheres the value given by Eq. \eqref{inflectionlimit} is $x_0/L=0.2537$. Thus, if we use Eq. \eqref{inflectionlimit} to estimate the inflection point, the relative error is $2\%$ and  $6\%$, respectively, for the two cases mentioned above. However, it worth to noting that the thickness of electrolyte does influence the position of inflection point. We show in the S.3 that Eq. \eqref{inflectionlimit} \textit{does} give good approximation on the inflection point as long as the thickness of the electrolyte is within a normal range ($10$ to $50 \mu m$).

\section{The effect of the critical current densities and operating temperatures on the degradation mechanisms}

According to results we show in section 4, in order to avoid the degradation in SOEC, a low current density and high operating temperature are expected. However, for the commercial concern, operating the cells at high current densities and low temperatures is preferred to maintain high efficiency and low production costs. Thus,
in the following, we focus on the critical condition, e.g., given an operating temperature ($T$), what is the biggest current
density ($i_{cr}$) we can run without having degradation at least due to the reasons mentioned in (C1), (C2) and (C3). 

According to the model we proposed in section 2, the critical current density ($i_{cr}$) (in magnitude) vs. operating temperatures (T) due to degradation mentioned in (C1) and (C2) are
\begin{align}
&\eta_O^{cr}=\frac{RT}{F}\sinh^{-1}\left(\frac{i_{cr}R_P^O F}{RT}\right)
=\frac{RT}{4F}\ln\left(\frac{P_{O_2}^{O-cr}}{P_{O_2}^{O, amb}}\right)
\label{A2}\end{align}
and
\begin{align}
\eta_H^{cr}&=\frac{RT}{F}\sinh^{-1}\left(\frac{i_{cr}R_P^H F}{RT}\right)\nonumber\\
&=-\frac{RT}{4F}\ln\left(\frac{P_{O_2}^{H-cr}}{P_{O_2}^{H-OCV}}\right)-\frac{RT}{2F}\ln\left(\frac{C_{V}^{H-cr}}{C_{V}^{H-OCV}}\right)+\frac{RT}{2F}\ln\left(\frac{C_{O^O}^{H-cr}}{C_{O^O}^{H-OCV}}\right)
\label{B2}\end{align}
respectively.

According to our estimate of the peak oxygen partial pressure at YSZ/GDC shown in Eq. \eqref{EstiPO2}, the critical value of current density ($i_{cr}$) (in magnitude) vs. operating temperatures (T) of having degradation due to (C3) mentioned above are obtained by
\begin{align}
P_{O_2}^{I-cr}= &P_{O_2}^{O,amb}\exp\left[-
\frac{16i_{cr}FL_{{GDC}}}{RT\sigma_h^{{GDC}}}\left(\frac{\sigma_h^{YSZ}}{\sigma_V^{YSZ}}-\frac{\sigma_h^{GDC}}{\sigma_V^{GDC}}\right)\sinh^{-1}\left(\frac{i_{cr}R_P^OF}{RT}\right)\right].
\label{C3}
\end{align}

Then, $i_{cr}$ for each mechanism mentioned above can be solved by Eq. \eqref{A2}, \eqref{B2} and \eqref{C3} for a given temperature $T$.

\begin{figure}[htpb]\centering
\includegraphics[scale=0.3]{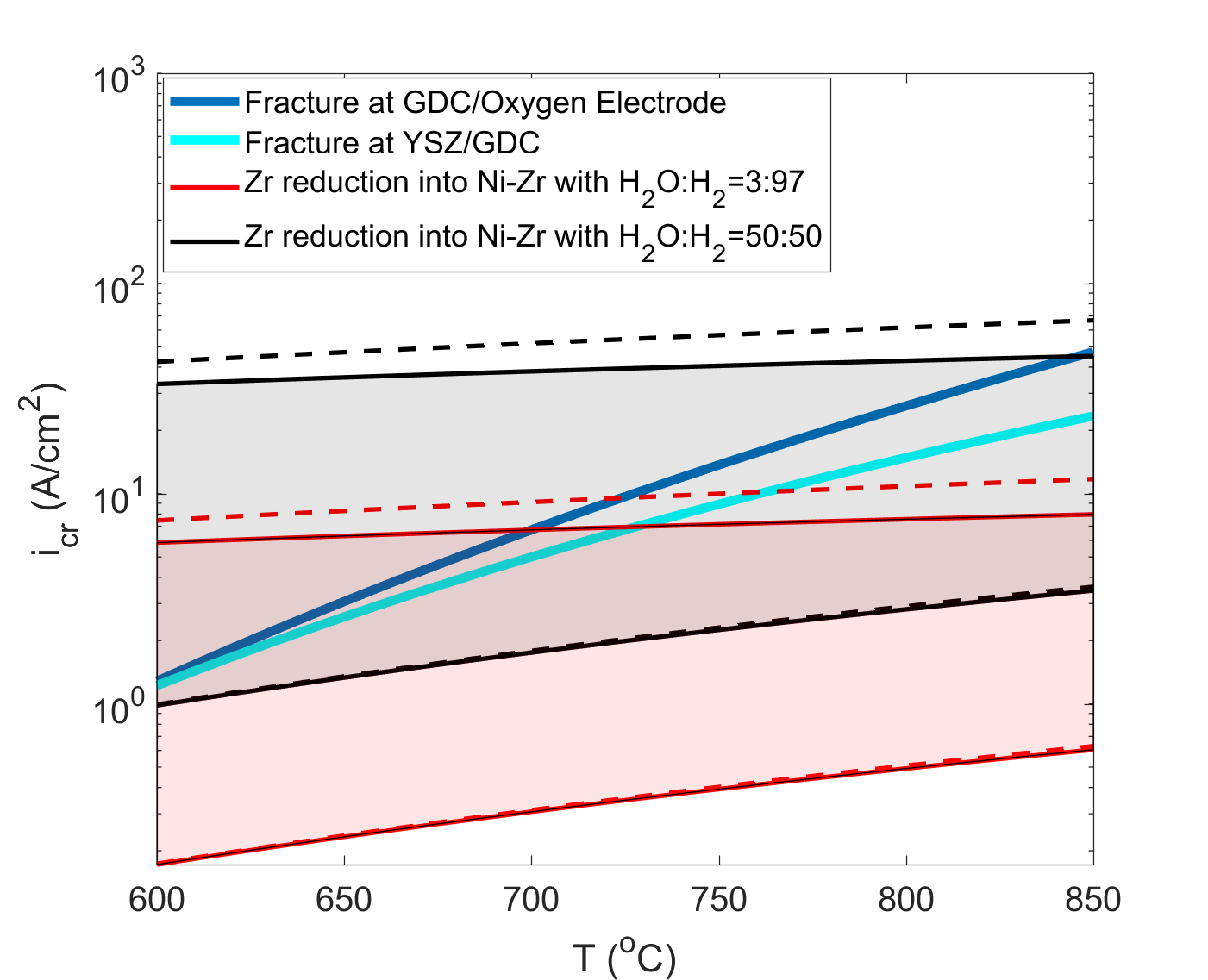}
\caption{Critical condition of having degradation mentioned in (C1), (C2) and (C3) in section  6. The red and black areas  count the uncertainty range of Gibbs energy of the corresponding reaction. Experimental data used to generate this figure is summarized in S.5. The dashed black lines and red lines are the upper bound and lower bound for the Zr reduction into Ni-Zr, when varying of molar concentration of oxygen vacancies does not take into account. }
\label{CRIT}
\end{figure}

For example, if we consider a full cell with structure in the following,
\begin{align*}
H_2:H_2O:Ni/YSZ:YSZ(16\mu m):GDC(4\mu m):STFC:Air
\end{align*}
we can get the critical conditions (current density $i_{cr}$ vs. operating temperature $T$) for degradation (C1), (C2) and (C3) obtained by Eq. \eqref{A2},\eqref{B2} and \eqref{C3} shown in Fig. \ref{CRIT}. For a given operating temperature, in order to avoid degradation introduced due to mechanisms (C1), (C2) and (C3), a current density lower than its value on dark blue line, blue line and red areas is preferred. Note that according to our discussion about mechanism (C2) in Section 2, the critical oxygen partial pressure of reducing $ZrO_{2}$ into Ni-Zr compound lies in a range, if taking the uncertainty range of Gibbs energy of the corresponding chemical reaction into account. That is why the critical current density at a temperature of getting reduction is a range rather than a value as shown in Fig. \ref{CRIT} (red area).
In particular, decreasing the concentration of $H_2$ at the fuel electrode increase the tolerances current density of getting degradation close to fuel electrode due to reduction of ZrO$_2$ significantly as shown in Fig. \ref{CRIT} (comparing red and black region). For the sake of comparison, we show the upper bound and lower bound for the onset of reduction in which the change of molar composition of oxygen vacancies is not taken into account (dashed red and black lines in Fig. \ref{CRIT}). Taking temperature at 600$^o$C for example, the upper bound of current density given by our model in which the changing of molar concentration is taken into account with fuel gas composition $3\%$ is $5.8A/cm^2$, while in the case molar concentration of oxygen vacancy is taken as a constant, the value is $7.4A/cm^2$. The relative difference is $27\%$. 

\section{Conclusions}
We examine the distribution of oxygen partial pressure in solid oxide electrolyser cells with  YSZ-GDC two-layer electrolytes of different widths and under different operating conditions.  Since we consider changes in the oxygen vacancy concentration, our results capture the oxygen potential distribution under highly reducing conditions at the hydrogen electrode.

An analytical estimate of the oxygen partial pressure at
the YSZ/GDC interface when a GDC barrier layer present is obtained and is in good agreement with the numerical results. 
The predictions are also in accord with experimental observations of fracture or void formation near
the GDC/YSZ interface. 
This result suggests that the peak oxygen partial pressure at the interface of YSZ/GDC  is present because of the differences in the conductivity ratios between YSZ and GDC, specifically when $\sigma_h^{YSZ}/\sigma_V^{YSZ}$ is smaller than $\sigma_h^{GDC}/\sigma_V^{GDC}$. Altering the electrolyte materials in order to affect their transport properties, e.g. changing the Gd doping level of GDC to increase the oxygen vacancy diffusivity should thereby suppress the formation of pores or cracks at the YSZ/GDC interface. The magnitude of the peak pressure can be reduced by reducing the GDC layer thickness and using  a lower polarization resistance oxygen electrode.  Changes in operating conditions, i.e.,  lowering the current density and increasing operating temperature can also reduce the peak pressure.

In addition, we provide an analytical estimate for the location of the inflection point in the spatial distribution of oxygen partial pressure for  YSZ8 with standard electrolyte thicknesses. The estimate shows that the position of inflection point depends mainly on the ratio of diffusivity of holes to electrons $D_h/D_e$, and the concentration of electrons at the interface of electrolyte and two electrodes. We show that other properties, such as the diffusivity of oxygen vacancies and thickness of electrolyte, also affects  the position of inflection point.   Using the analytical expressions for the oxygen partial pressures at the interface of YSZ and GDC, the GDC/Oxygen electrode and the YSZ/Hydrogen electrode for a given operating temperature, the critical current density below which various degradation phenomena will not occur can be easily connected with the   polarization resistances  measured in the experiments.

Our model is not only applicable to YSZ and GDC but  also  to other mixed ionic electronic conductors (MIECs) electrolyte materials, such as LSGM, scandia-stabilized zirconia (ScSZ). Our results can thus can be generalized to other single-layer or multi-layer electrolytes.
\section{Acknowledgment}
This research was supported by the Department of Energy under Grant DE-EE0008079.
\newpage

\end{document}